\documentclass[smallextended]{svjour3}
\usepackage{type1cm}
\usepackage[dvipdfm]{graphicx}
\usepackage{amsmath,amssymb,bm}

\def\beq{\begin{equation}}
\def\eeq{\end{equation}}
\def\nbeq{\begin{equation*}}
\def\neeq{\end{equation*}}
\def\<{\langle}
\def\>{\rangle}
\def\rt#1{\sqrt{\mathstrut #1}}
\def\Vec#1{\overrightarrow #1}

\renewcommand{\d}{\partial}

\begin{document}
\title{Microcanonical Analysis of Exactness of the Mean-Field Theory in Long-Range Interacting Systems}
\author{Takashi Mori}
\institute{Department of Physics, Graduate School of Science,The University of Tokyo, Bunkyo-ku, Tokyo 113-0033, Japan\\
\email{mori@spin.phys.s.u-tokyo.ac.jp}}
\titlerunning{Microcanonical Analysis of Exactness of the Mean-Field Theory in LRIS}
\maketitle

\begin{abstract}
Classical spin systems with nonadditive long-range interactions are studied in the microcanonical ensemble.
It is expected that the entropy of such a system is identical to that of the corresponding mean-field model,
which is called ``exactness of the mean-field theory''.
It is found out that this expectation is not necessarily true if the microcanonical ensemble is not equivalent to the canonical ensemble in the mean-field model.
Moreover, necessary and sufficient conditions for exactness of the mean-field theory are obtained.
These conditions are investigated for two concrete models, the $\alpha$-Potts model with annealed vacancies and the $\alpha$-Potts model with invisible states.
\end{abstract}

\section{Introduction}

Long-range interactions appear in several physical systems;
gravitational systems, non-neutral plasma, vortices in two-dimensional fluid, and so on~\cite{Bouchet2010,Campa2009,Lecture_notes2002,Les_Houches2008}.
Theoretically, long-range interacting systems (LRIS) show some peculiar features both in equilibrium and in nonequilibrium situations.
Ensemble inequivalence is one of those anomalous properties in LRIS, and has been studied for a long time~\cite{Barre2001,Bouchet2010,Campa2009,Leyvraz2002}.
In short-range interacting systems, several statistical ensembles, for instance the canonical ensemble and the microcanonical ensemble,
are equivalent in the sense that these thermodynamic potentials are related by Legendre transformation~\cite{Ruelle}.
However, in LRIS, two ensembles may exhibit different results, and this is called ensemble inequivalence
(see also~\cite{Bertalan2011ensemble} for spin glass systems).
For instance, it is known that the specific heat can be negative in the microcanonical ensemble, 
although it is always positive in the canonical ensemble~\cite{Thirring1970}.
In short-range interacting systems, the negative specific heat implies thermodynamical instability of the state.
Since the equilibrium state must be stable against any thermal fluctuation, the negative specific heat in equilibrium is forbidden.
It is nonadditivity that allows equilibrium states with negative specific heats in LRIS.
Nonadditivity means that the energy of two isolated subsystems is not equal to the total energy when they are combined~\cite{Les_Houches2008}.
Due to the nonadditivity, the usual argument on thermodynamic stability cannot be applied,
and the negative specific heat and the thermodynamic stability can coexist in LRIS.
LRIS also show anomalous features in relaxational dynamics such as the existence of the long-lived metastable states.
In order to understand the properties of those nonequilibrium quasi-stationary states,
dynamics of LRIS has been actively studied~\cite{Bachelard2011,Bouchet2010,Campa2009,Gupta2011,Pakter2011,Yamaguchi2004}.
These peculiar features in LRIS are ultimately related to nonadditivity.
Properties of systems without additivity have been less understood compared to those of short-range interacting systems.
Therefore, to understand many body systems with long-range interactions is an important and a challenging problem.

The so-called mean-field models (infinite range models) are extensively investigated because they are tractable models which display those unfamiliar characteristics of LRIS.
It is expected that general LRIS whose interaction decays as $1/r^{\alpha}$ with $0\leq\alpha<d$ in $d$-dimensional space
exhibit similar properties to the mean-field models.
Actually, many works support the {\it exactness of the mean-field theory} in LRIS;
that is, LRIS have {\it identical} equilibrium properties to the mean-field models~\cite{Barre2005,Campa2000,Cannas2000,Mori2010}.
However, if the system has some conserved quantities, the exactness of the mean-field theory does not necessarily hold.
In~\cite{Mori2010,Mori_instability}, it was shown that classical spin systems with long-range interactions 
have a temperature region where the mean-field theory is not exact
in the canonical ensemble with a fixed magnetization.

In this paper, we study classical spin systems with a long-range interaction. 
We consider systems whose magnetization is not fixed, but whose energy is conserved, that is, we treat them in the microcanonical ensemble.
Barr\'e discussed equilibrium states of the Ising model with long-range interactions in the microcanonical ensemble,
and argued that equilibrium properties of this system are independent of the power of the interaction $\alpha$ 
as long as $0\leq\alpha<d$~\cite{Barre2002microcanonical}.
Campa {\it et al.} calculated the partition function of the $n$-vevctor spin models and concluded that
the exactness of the mean-field theory holds in these systems both in the canonical ensemble and in the microcanonical ensemble~\cite{Campa2003}.
We will show that these results do not hold in general.
In the microcanonical ensemble, the exactness of the mean-field theory is not necessarily valid.
Necessary and sufficient conditions for the exactness of the mean-field theory in the microcanonical ensemble will be presented.
We will see that these conditions are fully described by mean-field quantities and one parameter $U_{\rm max}$.
The parameter $U_{\rm max}$ characterizes the interaction potential $K(\bm{r})$ and is easily calculated.
Therefore, we can judge the validity of the exactness of the mean-field theory only by analyzing the mean-field model.

\section{Model}
\label{sec:model}

We consider a system described by the following Hamiltonian:
\beq
H=-\frac{1}{2}\sum_{ij}^NK(\bm{r}_i-\bm{r}_j)\Vec{\sigma}_i^{\rm{T}}\bm{J}\Vec{\sigma}_j -\Vec{h}\cdot\sum_i^N\Vec{\sigma}_i.
\label{eq:H}
\eeq
We consider the $d$-dimensional cubic lattice, and
each lattice point at $\bm{r}_i\in [1,L]^d\cap \mathbb{Z}^d$ is associated with a spin variable $\sigma_i$.
The number of lattice points (or spins) is denoted by $N=L^d$.
Spin variable $\Vec{\sigma}$ is a general $q$-component vector whose norm is assumed to be finite.
We can easily generalize the interaction potential $K(\bm{r})$ to a matrix form, $\sum_{a,b}K_{ab}\sigma_i^a\sigma_j^b$,
where $a$ and $b$ are components of $\Vec{\sigma}$,
but we restrict the Hamiltonian to the simple form~(\ref{eq:H}) in order to make the discussion simple.
The coupling constant matrix $\bm{J}$ is assumed to be symmetric and positive-semidefinite.
The external field $\Vec{h}$ is a $q$-dimensional vector.
The interaction potential is long-ranged.
We refer to LRIS as systems with the following interaction potentials (i) or (ii):
\begin{itemize}
\item[(i)] {\bf non-additive limit}: 
the Kac potential is given by $$K(\bm{r})=\gamma^d\phi(\gamma\bm{r})>0,\quad \int_{r<\gamma L}\phi(\bm{r})d^dr<+\infty.$$
We take the limit $\gamma\rightarrow 0$, $L\rightarrow\infty$ with $\gamma L$ fixed ($L$ is the linear dimension of the system).
Additionally we assume the existence of a decreasing function $\psi(r)$ such that 
$|\nabla\phi(\bm{r})|\leq\psi(r)$ for all $r\leq \gamma L$ and there exists a positive constant $C$ such that
$\int_{\delta<r<\gamma L}\psi(\bm{r})d^dr<C/\delta$ for all $\delta>0$.
Existence of the function $\psi$ restricts the sudden change and the rapid oscillation of the potential $\phi$, and
the condition $\int_{\delta<r<\gamma L}\psi(\bm{r})d^dr<C/\delta$ means that $\psi(\bm{r})\lesssim 1/r^{d+1}$ in short distance.
These conditions are necessary to derive Eqs.~(\ref{eq:f_variation}) and (\ref{eq:free_functional}) discussed later.
Note that it includes power-law potentials $K(\bm{r})\sim 1/r^{\alpha}$, $0\leq\alpha<d$.
\item[(ii)] {\bf van der Waals limit}:
The potential $K(\bm{r})$ is given by $$K(\bm{r})=\gamma^d\phi(\gamma\bm{r})>0,\quad \int\phi(\bm{r})d^dr<+\infty,$$ 
and we take the van der Waals limit, that is $\gamma\rightarrow 0$ after $L\rightarrow\infty$.
In the van der Waals limit, the thermodynamic potential becomes independent of the original interaction form $\phi$.
Actually, in the van der Waals limit, the model is exactly described by the mean-field theory with the Maxwell construction~\cite{Lebowitz_Penrose1966}.
\end{itemize}

An important difference between (i) non-additive limit and (ii) van der Waals limit is that
the system is nonadditive in (i) but it is additive in (ii).
The interaction range is roughly given by $\gamma^{-1}$.
In the non-additive limit, $\gamma L$ is fixed at a constant, and therefore the interaction range is comparable with the system size.
In this case, the system is nonadditive.
On the other hand, in the van der Waals limit, $\gamma\rightarrow 0$ is taken after the limit of $L\rightarrow\infty$.
This means that the interaction range is much longer than the microscopic length like the lattice spacing but much shorter than the system size.
In this case, the system is additive and does not display anomalies such as negative specific heats.

In the case of (i), the interaction potential depends on the system size $L$.
Therefore, we consider the limit of large system size and the limit of weak coupling simultaneously in a proper manner. 
This scaling procedure is called Kac's prescription~\cite{Kac1963}.
We normalize the interaction potential as
\beq
\sum_i^NK(\bm{r}_i)=1.
\label{eq:Kac}
\eeq
We can safely take the thermodynamic limit under this normalization because the system has extensivity (but the system is still non-additive).

We refer to the mean-field model corresponding to (\ref{eq:H}) as
\begin{align}
H_{\rm MF}=-\frac{1}{2N}\sum_{i,j}\Vec{\sigma}_i^{\rm{T}}\bm{J}\Vec{\sigma}_j -\Vec{h}\cdot\sum_i\Vec{\sigma}_i
&=N\left(-\frac{1}{2}\Vec{m}^{\rm T}\bm{J}\Vec{m}-\Vec{h}\cdot\Vec{m}\right) \nonumber \\
&\equiv Nu_{\rm MF}(\Vec{m}),
\end{align}
where the vector $\Vec{m}\equiv \frac{1}{N}\sum_i\Vec{\sigma}_i$ is the magnetization
and $u_{\rm MF}(\Vec{m})$ is the energy per spin in the mean-field model.

It has been revealed that the mean-field theory is exact for the canonical ensemble without fixing the magnetization
but it is not for the canonical ensemble with restriction of a magnetization value in LRIS with periodic boundary conditions~\cite{Mori2010}.
Thus it depends on the specific ensemble whether the mean-field theory is exact or not in LRIS.
In the present paper, we examine whether the exactness of the mean-field theory holds or not for the microcanonical ensemble
with an unrestricted magnetization value in classical spin systems.

\section{Microcanonical entropy}

We consider a spin system with a fixed energy and without fixing the magnetization.
Such a situation is described by the microcanonical ensemble with an unrestricted magnetization value.

In the microcanonical ensemble, the natural thermodynamic potential is the entropy per spin $s(\varepsilon)$ which is a function of the energy per spin $\varepsilon$.
The entropy is defined by the Boltzmann formula:
\beq
s(\varepsilon)=\lim_{N\rightarrow\infty}\frac{1}{N}\log W(\varepsilon),
\label{eq:micro_s}
\eeq
where $W(\varepsilon)$ is the number of states with an energy per spin $\varepsilon$.
We have chosen the temperature unit so that the Boltzmann constant is unity.
In systems with $short$-range interactions, the entropy is obtained by the Legendre transformation of the canonical free energy $f(\beta)$,
\beq
s_{\rm can}(\varepsilon)=\inf_{\beta\ge 0}\left[\beta(\varepsilon-f(\beta))\right],
\label{eq:can_s}
\eeq
where $\beta=1/T$ is the inverse temperature.
However, in LRIS, the above formula does not necessarily hold, $s(\varepsilon)\neq s_{\rm can}(\varepsilon)$, due to ensemble inequivalence.
Therefore, we must calculate the entropy directly from the microcanonical ensemble.
Later, we will distinguish these two entropies, Eqs.~(\ref{eq:micro_s}) and (\ref{eq:can_s}).
The former is referred to as the microcanonical entropy, or more simply, the entropy. 
The latter is called the canonical entropy.

The free energy $f(\beta)$ in the canonical ensemble is expressed as 
\beq
\beta f(\beta)=\inf_{\varepsilon}[\beta\varepsilon-s(\varepsilon)]=s^*(\beta),
\label{eq:Legendre}
\eeq
where the superscript $*$ means the Legendre transformation.
The canonical entropy (\ref{eq:can_s}) is thus expressed as $s_{\rm can}(\varepsilon)=s^{**}(\varepsilon)$,
namely, it is obtained by applying the Legendre transformation twice on $s(\varepsilon)$.
It is well known that $s^{**}(\varepsilon)$ is the concave envelope of $s(\varepsilon)$ 
(here the ``Legendre transformation'' of a concave function $f(x)$ is implemented by $f^*(y)=\inf_{x}[xy-f(x)]$).
The concave envelope of $s(\varepsilon)$ is defined as the smallest concave function greater than or equal to $s(\varepsilon)$ for all $\varepsilon$.
Therefore, the following relation
\beq
s_{\rm can}(\varepsilon)=s^{**}(\varepsilon)\geq s(\varepsilon)
\eeq
follows. The canonical entropy gives the upper bound of the microcanonical entropy.
Since the mean-field theory is exact in the canonical ensemble, that is, $f(\beta)=f_{\rm MF}(\beta)$ in LRIS,
$s_{\rm can}(\varepsilon)=s_{\rm MF}^{**}(\varepsilon)$ holds.
Therefore, $s(\varepsilon)\leq s_{\rm MF}^{**}(\varepsilon)$, 
that is, the microcanonical entropy is bounded above by the concave envelope of the mean-field entropy.

On the other hand, the microcanonical entropy of the mean-field model $s_{\rm MF}(\varepsilon)$ gives the lower bound of $s(\varepsilon)$ in general.
To show it, let us consider the following thermodynamic process.
We consider a Hamiltonian $H(\lambda)=(1-\lambda)H_{\rm MF}+\lambda H_1$.
Here $H_1=-\frac{1}{2}\sum_{ij}J_{ij}\sigma_i\sigma_j$ is the Hamiltonian of the system of interest ($J_{ij}$ is arbitrary)
and $H_{\rm MF}=-\frac{J}{2N}\sum_{ij}\sigma_i\sigma_j$ is the Hamiltonian of the corresponding mean-field model.
The parameter $J$ is set as $J=\sum_{ij}J_{ij}/N$.
Initially, we set $\lambda=0$ and prepare the microcanonical ensemble of the Hamiltonian $H_{\rm MF}$ with the energy per spin $\varepsilon$,
which is the initial state of the process.
Then, we isolate the system from the environment 
and change the parameter $\lambda(t)$ from $\lambda(0)=0$ at time $t=0$ to $\lambda(\tau)=1$ at time $t=\tau$.
Since this process is an adiabatic process, the entropy of the final state is greater than or equal to that of the initial state 
regardless of the protocol $\lambda(t)$ of changing $\lambda$.
Namely, $s(\< H\>_{\rm fin}/N)\geq s_{\rm MF}(\varepsilon)$.
The expectation value of the energy of the final state is denoted by $\< H\>_{\rm fin}$.
If we change the parameter $\lambda$ suddenly from 0 to 1 ($\tau$ is infinitesimal),
the expectation value of the energy is $\< H\>_{\rm fin}=\< H_1\>_{\varepsilon}^{\rm MF}$.
The symbol $\<\cdot\>_{\varepsilon}^{\rm MF}$ denotes the microcanonical average 
under the mean-field Hamiltonian $H_{\rm MF}$ and the energy $\varepsilon$ per spin.
It follows that
\begin{align}
\< H_1\>_{\varepsilon}^{\rm MF}&=-\frac{1}{2}\sum_{ij}J_{ij}\<\sigma_i\sigma_j\>_{\varepsilon}^{\rm MF} \nonumber \\
&=-\frac{1}{2}\sum_{ij}J_{ij}\<\sigma_i\>_{\varepsilon}^{\rm MF}\<\sigma_j\>_{\varepsilon}^{\rm MF} \nonumber \\
&=-\frac{1}{2}m^2\sum_{ij}J_{ij} \nonumber \\
&=-\frac{NJ}{2}m^2=\< H_{\rm MF}\>_{\varepsilon}^{\rm MF}=N\varepsilon,
\end{align}
where $m=\<\sum_i\sigma_i\>_{\varepsilon}^{\rm MF}/N$.
Thus $s(\varepsilon)\geq s_{\rm MF}(\varepsilon)$ is derived.
The same argument is applicable to more general Hamiltonians~(\ref{eq:H}).

The coarse graining of the spin configuration is expected to be performed without errors in LRIS 
because short-scale structure of the local magnetization does not contribute to the free energy per spin.
In LRIS, indeed, it is known that the free energy in the canonical ensemble is given by the following expression~\cite{Mori2010}:
\beq
f(\beta)=\min_{\Vec{m}(\bm{x})}{\cal F}(\beta,\{\Vec{m}(\bm{x})\})
\label{eq:f_variation}
\eeq
with the free energy functional ${\cal F}$ which is an analytic functional of $\beta$ and $\Vec{m}(\bm{x})$.
The $\Vec{m}(\bm{x})$ is the coarse-grained magnetization at a point $\bm{x}$.
For the Hamiltonian~(\ref{eq:H}), the free energy functional is given by
\begin{align}
{\cal F}(\beta,\{ \Vec{m}(\bm{x})\})=&-\frac{1}{2}\int_{C_d}d^dx\int_{C_d}d^dyU(\bm{x}-\bm{y})\Vec{m}^{\rm T}(\bm{x})\bm{J}\Vec{m}(\bm{y})
-\Vec{h}\cdot\int_{C_d}d^dx\Vec{m}(\bm{x}) \nonumber \\
&-T\int_{C_d}d^dx\omega(\Vec{m}(\bm{x}))\nonumber \\
\equiv & \ {\cal U}[\Vec{m}(\bm{x})]-T\int_{C_d}d^dx\omega(\Vec{m}(\bm{x})),
\label{eq:free_functional}
\end{align}
where $C_d$ is the $d$-dimensional unit cube.
The scaled coordinate $\bm{x}\in C_d$ is related to the original position vector $\bm{r}_i$ by $\bm{x}=\bm{r}_i/L$.  
The function $\omega(\Vec{m})$ is defined as
\begin{align}
\omega(\Vec{m})=\lim_{N\rightarrow\infty}\frac{1}{N}
&\log\Biggl(\text{the number of states with a fixed magnetization } \nonumber \\
&\left.\Vec{m}=\frac{1}{N}\sum_i^N\Vec{\sigma}_i\right),
\end{align}
which is referred to as the ``rate function'' in this paper.
The scaled potential $U(\bm{x})$ is given by
\beq
U(\bm{x})\equiv \lim_{L\rightarrow\infty}L^dK(L\bm{x}).
\eeq
In the case of (i) non-additive limit, because of $\gamma L=1$, the scaled potential is nothing but $\phi(\bm{x})$.
In the case of (ii) van der Waals limit, $U(\bm{x})=\delta(\bm{x})$.
The free energy can be expressed by the free energy functional in both cases,
but the interaction potential $U(\bm{x})$ is non-local even after the coarse-graining in the non-additive limit.

In order to derive Eq.~(\ref{eq:f_variation}) in the non-additive limit,
the conditions on the non-additive limit discussed in Sec.~\ref{sec:model}, e.g. the presence of $\psi(r)$, is necessary.
In this paper, this derivation is not discussed.
For the derivation, see the reference~\cite{Mori_instability}.

The minimum value of the free energy functional for $\Vec{m}(\bm{x})$ gives the free energy in the canonical ensemble.
If the magnetization is uniform, $\Vec{m}(\bm{x})=\Vec{m}$, the free energy functional is the same as the mean-field free energy with a fixed magnetization:
$$f_{\rm MF}(\beta,\Vec{m})={\cal F}(\beta,\{\Vec{m}(\bm{x})=\Vec{m}\}).$$
Namely, if the magnetization is uniform in the space, this state is described by the mean-field theory.
On the other hand, if inhomogeneity appears in the equilibrium state, it is a sign of violation of the exactness of the mean-field theory.

In LRIS, the original Hamiltonian (\ref{eq:H}) defined on a lattice can be approximated by the ``energy functional'' ${\cal U}[\Vec{m}(\bm{x})]$.
Indeed, we can show that $H=N{\cal U}[\Vec{m}(\bm{x})]+o(N)$ for LRIS (see \cite{Mori_instability}).
Therefore, the microcanonical entropy is expressed as
\beq
s(\varepsilon)=\sup_{\Vec{m}(\bm{x})}\left\{\int_{C_d}\omega(\Vec{m}(\bm{x}))d^dx: 
\varepsilon={\cal U}[\Vec{m}(\bm{x})]\right\}.
\label{eq:entropy}
\eeq
The notation $\sup[A:B]$ means that the maximum value of $A$ under the condition $B$.
The entropy $s(\varepsilon)$ is not obtained by the Legendre transformation~(\ref{eq:can_s}),
but a similar formula
\beq
s(\varepsilon)=\sup_{\Vec{m}(\bm{x})}\inf_{\beta\geq 0}\left\{\beta\left[\varepsilon-{\cal F}(\beta,\{\Vec{m}(\bm{x})\})\right]\right\}
\label{eq:s_Legendre}
\eeq
holds.
By substituting Eq.~(\ref{eq:free_functional}) into Eq.~(\ref{eq:s_Legendre}), we obtain
\begin{align}
s(\varepsilon)&=\sup_{\Vec{m}(\bm{x})}\inf_{\beta>0}\left[\beta\left(\varepsilon-{\cal U}[\Vec{m}(\bm{x})]\right)
+\int_{C_d}\omega(\Vec{m}(\bm{x}))d^dx\right] \nonumber \\
&=\max\left[\sup_{\Vec{m}(\bm{x})}\inf_{\beta>0}\left[\beta\left(\varepsilon-{\cal U}[\Vec{m}(\bm{x})]\right)
+\int_{C_d}\omega(\Vec{m}(\bm{x}))d^dx : \varepsilon\geq {\cal U}[\Vec{m}(\bm{x})]\right],\right. \nonumber \\
&\left. \qquad\sup_{\Vec{m}(\bm{x})}\inf_{\beta>0}\left[\beta\left(\varepsilon-{\cal U}[\Vec{m}(\bm{x})]\right)
+\int_{C_d}\omega(\Vec{m}(\bm{x}))d^dx : \varepsilon<{\cal U}[\Vec{m}(\bm{x})]\right]\right]
\end{align}
In the last equality, we split the configurations $\{\Vec{m}(\bm{x})\}$ into those of $\varepsilon\geq{\cal U}[\Vec{m}(\bm{x})]$ and
$\varepsilon<{\cal U}[\Vec{m}(\bm{x})]$
In the former, the minimum for $\beta$ is realized at $\beta=0$.
In the latter, there exists no minimum point for $\beta$.
Therefore, it is confirmed that the expression~(\ref{eq:s_Legendre}) is equivalent to
\beq
s(\varepsilon)=\sup_{\Vec{m}(\bm{x})}\left\{\int_{C_d}\omega(\Vec{m}(\bm{x}))d^dx: 
\varepsilon\geq{\cal U}[\Vec{m}(\bm{x})]\right\}.
\label{eq:entropy2}
\eeq
The expression~(\ref{eq:entropy2}) is equivalent to Eq.~(\ref{eq:entropy}) in the thermodynamic limit 
as long as negative temperatures are ruled out.

If we exchange sup and inf in Eq.~(\ref{eq:s_Legendre}), it agrees to the canonical entropy (see Eq.~(\ref{eq:can_s})),
$$\inf_{\beta\geq 0}\sup_{\Vec{m}(\bm{x})}\left\{\beta\left[\varepsilon-{\cal F}(\beta,\{\Vec{m}(\bm{x})\})\right]\right\}
=\inf_{\beta>0}[\beta(\varepsilon-f(\beta))]=s_{\rm can}(\varepsilon).$$
Again the inequality $s(\varepsilon)\leq s_{\rm can}(\varepsilon)$ is obtained immediately from (\ref{eq:s_Legendre}) 
because $\sup\inf(\cdot)\leq\inf\sup(\cdot)$.
On the other hand, if we restrict $\Vec{m}(\bm{x})$ to uniform functions of $\bm{x}$ in Eq.~(\ref{eq:entropy}),
\beq
s(\varepsilon)\geq\sup_{\Vec{m}}\left[\omega(\Vec{m}): \varepsilon=-\frac{1}{2}\Vec{m}^{\rm T}\bm{J}\Vec{m}-\Vec{h}\cdot\Vec{m}\right]
=s_{\rm MF}(\varepsilon),
\eeq
which was already derived by the thermodynamic argument.

In conclusion, we found that the entropy satisfies the inequality
\beq
s_{\rm MF}(\varepsilon)\leq s(\varepsilon)\leq s^{**}_{\rm MF}(\varepsilon).
\label{eq:entropy_ineq}
\eeq
If the mean-field entropy is a concave function of $\varepsilon$,
it turns out from Eq.~(\ref{eq:entropy_ineq}) that the exactness of the mean-field theory holds also for the microcanonical ensemble, 
$s(\varepsilon)=s_{\rm MF}(\varepsilon)$.
In other words, if the microcanonical ensemble is equivalent to the canonical ensemble in the mean-field model,
then these two ensembles remain equivalent even if the interaction potential is replaced by a slowly decaying function.
In Ref.~\cite{Barre2002microcanonical}, the long-range Ising model was studied in the microcanonical ensemble and
it was shown that the exactness of the mean-field theory holds in this model if the ferromagnetic interaction decays slower than $1/r^{d}$.
Campa {\it et al.}~\cite{Campa2003} also obtained the same result for a family of $n$-vector spin models.
It should be noted that the mean-field versions of these models satisfy $s_{\rm MF}(\varepsilon)=s_{\rm MF}^{**}(\varepsilon)$,
and the inequality (\ref{eq:entropy_ineq}) explains the results of Ref.~\cite{Barre2002microcanonical,Campa2003}.

However, it is known that the microcanonical entropy is not necessarily concave in LRIS;
negative specific heats are observed in some models.
In those models, we cannot conclude from Eq.~(\ref{eq:entropy_ineq}) that the exactness of the mean-field theory in LRIS holds in the microcanonical ensemble.

\subsection*{\bf{Remark on the van der Waals limit}.}
It is verified from Eqs.~(\ref{eq:entropy}) and (\ref{eq:U_vdw}) that $s(\varepsilon)=s^{**}_{\rm MF}(\varepsilon)$ in the van der Waals limit.
Because $U(\bm{x})=\delta(\bm{x})$ in the van der Waals limit, the energy functional is of the form,
\beq
{\cal U}[\Vec{m}(\bm{x})]=\int_{C_d}u_{\rm MF}(\Vec{m}(\bm{x}))d^dx.
\label{eq:U_vdw}
\eeq
Substituting Eq.~(\ref{eq:U_vdw}) to Eq.~(\ref{eq:entropy}), we obtain
\begin{align}
s(\varepsilon)&=\sup_{\Vec{m}(\bm{x})}\left\{\int_{C_d}\omega(\Vec{m}(\bm{x}))d^dx:\varepsilon=\int_{C_d}u_{\rm MF}(\Vec{m}(\bm{x}))d^dx\right\}
\nonumber \\
&=\sup_{\varepsilon(\bm{x})}\left\{\int_{C_d}\sup_{\Vec{m}(\bm{x})}\left[\omega(\Vec{m}(\bm{x})):u_{\rm MF}(\Vec{m}(\bm{x}))=\varepsilon(\bm{x})\right]d^dx:\int_{C_d}\varepsilon(\bm{x})d^dx=\varepsilon\right\} 
\nonumber \\
&=\sup_{\varepsilon(\bm{x})}\left[\int_{C_d}s_{\rm MF}(\varepsilon(\bm{x}))d^dx:\int_{C_d}\varepsilon(\bm{x})d^dx=\varepsilon\right]
\nonumber \\
&=s_{\rm MF}^{**}(\varepsilon).
\label{eq:s_vdw}
\end{align}
Last equality is derived by showing that the function 
$$\sup_{\varepsilon(\bm{x})}\left[\int_{C_d}s_{\rm MF}(\varepsilon(\bm{x}))d^dx:\int_{C_d}\varepsilon(\bm{x})d^dx=\varepsilon\right]$$
is a concave function of $\varepsilon$ (Proof of the concavity is essentially the same as Appendix. C of \cite{Mori_instability}.).
The last equality of Eq.~(\ref{eq:s_vdw}) is obtained by combining this fact with the inequality 
$s_{\rm MF}(\varepsilon)\leq s(\varepsilon)\leq s_{\rm MF}^{**}(\varepsilon)$,
since $s_{\rm MF}^{**}(\varepsilon)$ is the smallest concave function greater than or equal to $s_{\rm MF}(\varepsilon)$.

Equation~(\ref{eq:s_vdw}) means that the microcanonical entropy is obtained by the mean-field theory with the Maxwell construction~\cite{Campa2009}.
This result was well known one obtained by Lebowitz and Penrose~\cite{Lebowitz_Penrose1966}.
In later sections, we focus on the case of (ii) non-additive limit.

\section{Local stability of the uniform solutions}

We study the local stability of uniform states by the Lagrange multiplier method.
Hereafter, we impose the periodic boundary condition.
First, for convenience, we rewrite the energy functional by the Fourier modes,
\beq
{\cal U}[\Vec{m}(\bm{x})]=-\frac{1}{2}\sum_{\bm{n}\in\mathbb{Z}^d}U_{\bm{n}}\Vec{m}^{\rm T}_{\bm{n}}\bm{J}\Vec{m}_{-\bm{n}}
-\Vec{h}\cdot\Vec{m}_0,
\eeq
where
\beq
\left\{
\begin{split}
\Vec{m}_{\bm{n}}&=\int_{C_d}d^dx \Vec{m}(\bm{x})e^{2\pi i\bm{n}\cdot\bm{x}}, \\
U_{\bm{n}}&=\int_{C_d}d^dx U(\bm{x})\cos(2\pi\bm{n}\cdot\bm{x}).
\end{split}
\right.
\eeq
Here, $U_{\bm{n}}$ is called the interaction eigenvalue.
The maximum interaction eigenvalue except for $\bm{n}\neq 0$ is denoted by $U_{\rm max}$, that is,
\beq
U_{\rm max}\equiv \max_{\bm{n}\neq 0}U_{\bm{n}}.
\eeq
This quantity plays important roles later.

As the total energy is fixed, we search the extremum of the Lagrange function
\beq
L(\{\Vec{m}_{\bm{n}}\},\beta)=\int_{C_d}\omega(\Vec{m}(\bm{x}))d^dx+\beta(\varepsilon-{\cal U}[\Vec{m}(\bm{x})]),
\eeq
where $\beta$ is a Lagrange multiplier.
Let us consider the uniform states, $\Vec{m}(\bm{x})=\Vec{m}$.
Then the extreme condition is obtained by performing the functional derivative $\delta/\delta \Vec{m}(\bm{x})$ and putting $\Vec{m}(\bm{x})=\Vec{m}$,
\beq
\frac{\d}{\d \Vec{m}}\omega(\Vec{m})+\beta(\bm{J}\Vec{m}+\Vec{h})=0
\label{eq:extreme1}
\eeq
and
\beq
\varepsilon=-\frac{1}{2}\Vec{m}^{\rm T}\bm{J}\Vec{m}-\Vec{h}\cdot\Vec{m}.
\label{eq:extreme2}
\eeq
If $\Vec{m}^*$ denotes the magnetization which satisfies Eqs.~(\ref{eq:extreme1}) and (\ref{eq:extreme2}),
the following relation holds:
\beq
\beta(\bm{J}\Vec{m}^*+\Vec{h})=-\frac{\d}{\d \Vec{m}^*}\omega(\Vec{m}^*).
\label{eq:local1}
\eeq
If we regard $\Vec{m}^*$ as a function of $\varepsilon$, we obtain $\omega(\Vec{m}^*(\varepsilon))=s_{\rm MF}(\varepsilon)$ and
\beq
\frac{\d}{\d\Vec{m}^*}=\frac{\d\varepsilon}{\d\Vec{m}^*}\frac{\d}{\d\varepsilon}=-(\bm{J}\Vec{m}^*+\Vec{h})\frac{\d}{\d\varepsilon}.
\label{eq:local2}
\eeq
Therefore, we have
\beq
\beta=\frac{\d s_{\rm MF}(\varepsilon)}{\d\varepsilon}\equiv\beta_{\rm MF}(\varepsilon).
\eeq
This is nothing but the microcanonical temperature of the mean-field model.
One of the obtained solutions $\beta$ and $\Vec{m}^*$ corresponds to an equilibrium state of the mean-field model.

Next, we consider local stability of an equilibrium state $\Vec{m}^*(\varepsilon)$ of the mean-field model.
If there is  some small fluctuation which increases the entropy $\int_{C_d}d^dx\omega(\Vec{m}(\bm{x}))$, it is said that this state is locally unstable.
The stability problem under constraints is solved by analyzing the sign of the minor determinant of the bordered Hessian matrix 
(for example, see \cite{Magnus_matrix} and Appendix~\ref{sec:BH}).
It is found that the bordered Hessian matrix ${\cal H}$ is block diagonalized into the space with different wavenumbers $\{\pm\bm{n}\}$,
that is, ${\cal H}={\rm diag}({\cal H}_{\pm\bm{n}})_{\bm{n}\in\mathbb{Z}^d,n_1\geq 0}$.
The block of $\bm{n}=0$ is
\beq
{\cal H}_0=
\begin{pmatrix}
0&-\left(\frac{\d {\cal U}[\Vec{m}(\bm{x})]}{\d\Vec{m}_0}\right)^{\rm T} \\
-\frac{\d{\cal U}[\Vec{m}(\bm{x})]}{\d\Vec{m}_0}&\frac{\d^2L}{\d\Vec{m}_0\d\Vec{m}_0}
\end{pmatrix}.
\eeq
Here $\d^2L/(\d\Vec{m}_0\d\Vec{m}_0)$ is a $q\times q$ matrix whose matrix elements are given by
$$\left(\frac{\d^2L}{\d\Vec{m}_0\d\Vec{m}_0}\right)_{ab}=\frac{\d^2L}{\d m_0^a\d m_0^b}.$$
Since the uniform state is stable in the mean-field model, determinant of $(-1)^q{\cal H}_0$ is positive.
The block of $\pm\bm{n}\neq 0$ is
\beq
{\cal H}_{\pm\bm{n}}=\frac{\d^2L(\{\Vec{m}_{\bm{n}}\},\beta)}{\d\Vec{m}_{\bm{n}}\d\Vec{m}_{-\bm{n}}}
=\frac{\d^2\omega(\Vec{m})}{\d\Vec{m}\d\Vec{m}}+\beta U_{\bm{n}}\bm{J}.
\label{eq:BH_n}
\eeq
Because the bordered Hessian matrix is block diagonalized into spaces with different wavenumber $\pm\bm{n}$,
the problem is reduced to an eigenvalue problem of each block ${\cal H}_{\pm\bm{n}}$.
We can judge the stability of the equilibrium state of the mean-field model by analyzing only the sign of the maximum eigenvalue of $\{{\cal H}_{\pm\bm{n}}\}$.
Since the matrix $\bm{J}$ is assumed to be positive semidefinite, the maximum eigenvalue of RHS of Eq.~(\ref{eq:BH_n}) is an increasing function of $U_{\bm{n}}$.
Therefore, the maximum eigenvalue of
\beq
{\cal H}_{\rm max}\equiv\frac{\d^2\omega(\Vec{m})}{\d\Vec{m}\d\Vec{m}}+\beta U_{\rm max}\bm{J}
\label{eq:BH_max}
\eeq
determines the stability.
{\it The uniform state is locally unstable and the mean-field model becomes non-exact
if the maximum eigenvalue of the matrix ${\cal H}_{\rm max}$ is positive.
Contrarily, if the maximum eigenvalue of ${\cal H}_{\rm max}$ is negative,
the uniform state described by the mean-field model is at least locally stable}.
This is a necessary condition for the exactness of the mean-field theory.

Finally, we mention the relation between the bordered Hessian matrix and the microcanonical entropy.
From Eqs.~(\ref{eq:local1}) and (\ref{eq:local2}), we obtain
\begin{align}
\frac{\d^2\omega(\Vec{m}^*)}{\d\Vec{m}^*\d\Vec{m}^*}
&=\frac{\d}{\d\Vec{m}^*}\left[-\beta_{\rm MF}(\bm{J}\Vec{m}^*+\Vec{h})\right] \nonumber \\
&=-\beta_{\rm MF}\bm{J}+(\bm{J}\Vec{m}^*+\Vec{h})(\bm{J}\Vec{m}^*+\Vec{h})^{\rm T}\frac{\d\beta_{\rm MF}}{\d\varepsilon}.
\end{align}
Therefore, each block of the Hessian matrix (\ref{eq:BH_n}) is expressed as
\beq
{\cal H}_{\pm\bm{n}}=-(1-U_{\bm{n}})\beta_{\rm MF}(\varepsilon)\bm{J}+\bm{W}\frac{\d^2s_{\rm MF}(\varepsilon)}{\d\varepsilon^2},
\label{eq:BH_n2}
\eeq
where we defined $\bm{W}\equiv(\bm{J}\Vec{m}^*+\Vec{h})(\bm{J}\Vec{m}^*+\Vec{h})^{\rm T}$.

The first term of RHS of Eq.~(\ref{eq:BH_n2}) is a negative-semidefinite matrix.
Because the matrix $\bm{W}$ is positive-semidefinite, the second term of Eq.~(\ref{eq:BH_n2}) is also negative-semidefinite 
if $\d^2s_{\rm MF}(\varepsilon)/\d\varepsilon^2<0$.
Therefore, the maximum eigenvalue of (\ref{eq:BH_n}) must be negative in the energy region of $\d^2s_{\rm MF}(\varepsilon)/\d\varepsilon^2<0$.
Namely, the uniform state described by the mean-field theory is locally stable if the specific heat of this state is positive.
On the other hand, if the specific heat is negative, that is $\d^2s_{\rm MF}(\varepsilon)/\d\varepsilon^2>0$, 
the maximum eigenvalue may be positive.
Actually, in the van der Waals limit ($U_{\bm{n}}=1$ $\forall \bm{n}\in \mathbb{Z}^d$), 
${\cal H}_{\pm\bm{n}}=\bm{W}\d^2s_{\rm MF}(\varepsilon)/\d\varepsilon^2$ and
all the homogeneous states with negative specific heat are unstable.
In the non-additive limit ($U_{\rm max}<1$), two energy regions appear in general;
a region where homogeneous states described by the mean-field theory are locally stable,
and the other region where they are locally unstable.

\section{Upper bound of the microcanonical entropy}

In the previous section, we derived a {\it necessary} condition for the exactness of the mean-field theory.
In this section, we investigate a {\it sufficient} condition by evaluating an upper bound of the microcanonical entropy.

From now on, we derive the following upper bound of the microcanonical entropy:
\beq
s(\varepsilon)\leq \sup_{\varepsilon'\geq \varepsilon}\left[\frac{\beta_{\rm conv}}{U_{\rm max}}(\varepsilon-\varepsilon')+s_{\rm MF}(\varepsilon')\right],
\label{eq:s_upper}
\eeq
where $\beta_{\rm conv}$ is the inverse temperature where the convexity of the mean-field free energy is lost,
\beq
\beta_{\rm conv}\equiv \sup\left\{\beta\geq 0:f_{\rm MF}(\beta,\Vec{m})=f_{\rm MF}^{**}(\beta,\Vec{m}) \forall \Vec{m}\right\},
\eeq
where $f_{\rm MF}^{**}(\beta,\Vec{m})$ is the convex envelope of $f_{\rm MF}(\beta,\Vec{m})$ with respect to $\Vec{m}$.
This relation will be proved in Sec.~\ref{sec:derivation}.

From this relation, we conclude the following statement, which will be useful to check whether $s(\varepsilon)=s_{\rm MF}(\varepsilon)$ holds;
{\it If the function in the bracket of Eq.~(\ref{eq:s_upper}) takes the maximum value at $\varepsilon'=\varepsilon$,
we obtain $s(\varepsilon)\leq s_{\rm MF}(\varepsilon)$ as an upper bound.
In this case, combined with $s(\varepsilon)\geq s_{\rm MF}(\varepsilon)$, we find that the mean-field theory gives the exact microcanonical entropy, 
$s(\varepsilon)=s_{\rm MF}(\varepsilon)$.}

\subsection{Graphical meaning of the upper bound}

\begin{figure}[tbh]
\begin{center}
\begin{tabular}{cc}
(a)&(b)\\
\includegraphics[clip,width=6cm]{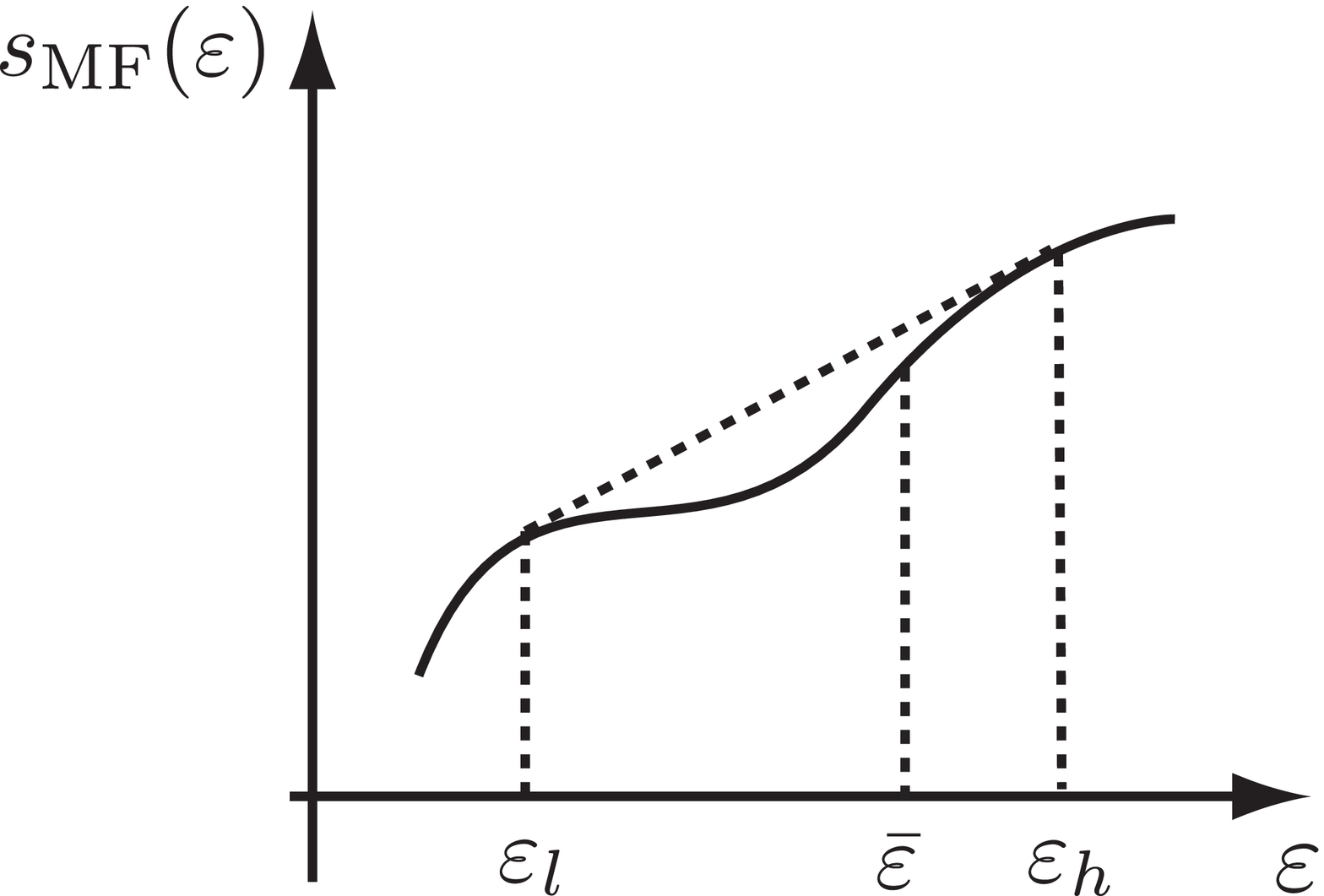}&
\includegraphics[clip,width=5cm]{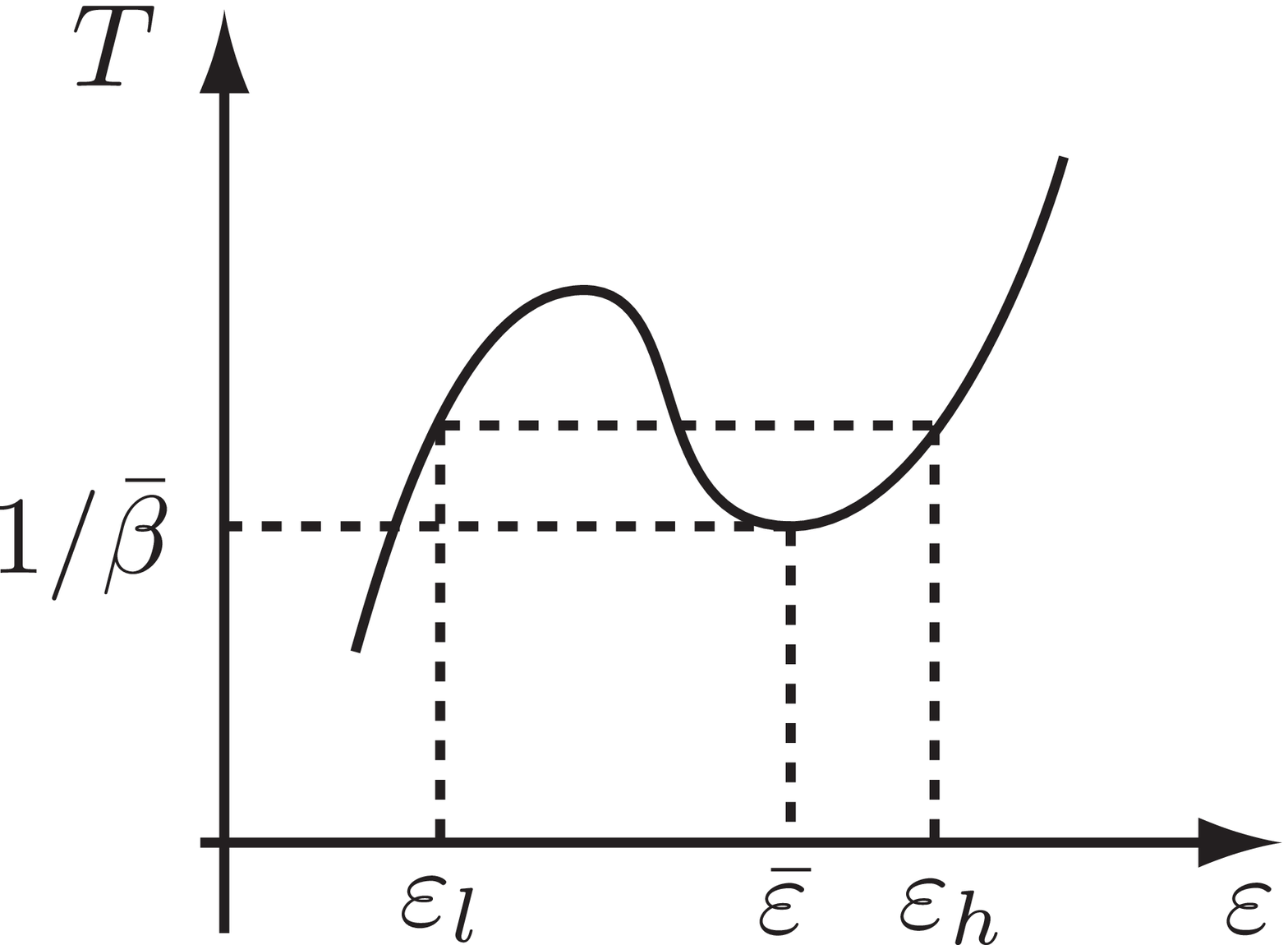}
\end{tabular}
\caption{Outlines of (a) the mean-field entropy and (b) the mean-field temperature as functions of the energy.}
\label{fig:illustration}
\end{center}
\end{figure}

It helps our understanding to graphically represent the upper bound (\ref{eq:s_upper}).
In Fig.~\ref{fig:illustration}(a), we depict a typical shape of the mean-field entropy $s_{\rm MF}(\varepsilon)$.
Here, we assume that $s_{\rm MF}(\varepsilon)\neq s_{\rm MF}^{**}(\varepsilon)$ in the region $\varepsilon_l<\varepsilon<\varepsilon_h$,
and focus on this region because $s(\varepsilon)=s_{\rm MF}(\varepsilon)$ holds trivially outside of this region due to the inequality~(\ref{eq:entropy_ineq}).
In Fig.~\ref{fig:illustration}~(a),
$\bar{\varepsilon}$ is the inflection point of $s_{\rm MF}(\varepsilon)$, 
and thus $$\frac{\d s_{\rm MF}(\varepsilon)}{\d\varepsilon}\equiv\beta_{\rm MF}(\varepsilon)\leq\beta_{\rm MF}(\bar{\varepsilon})\equiv\bar{\beta}.$$
More generally, we define 
\beq
\bar{\beta}\equiv\max_{\varepsilon_l\leq\varepsilon\leq\varepsilon_h}\beta_{\rm MF}(\varepsilon).
\eeq

From Eqs.~(\ref{eq:s_upper}) and (\ref{eq:entropy_ineq}), if
\beq
\frac{\beta_{\rm conv}}{U_{\rm max}}(\varepsilon-\varepsilon')+s_{\rm MF}(\varepsilon')\leq s_{\rm MF}(\varepsilon)
\label{eq:graphical1}
\eeq
for all $\varepsilon'\geq\varepsilon$, the mean-field theory is exact, $s(\varepsilon)=s_{\rm MF}(\varepsilon)$.
Inequality~(\ref{eq:graphical1}) is rewritten as
\beq
s_{\rm MF}(\varepsilon')\leq s_{\rm MF}(\varepsilon)+\frac{\beta_{\rm conv}}{U_{\rm max}}(\varepsilon'-\varepsilon).
\label{eq:graphical2}
\eeq
If we depict the graph of the RHS of Eq.~(\ref{eq:graphical2}) as a function of $\varepsilon'$,
it is expressed as a straight line of slope $\beta_{\rm conv}/U_{\rm max}$ which passes through a point $(\varepsilon,s_{\rm MF}(\varepsilon))$.
Therefore, Eq.~(\ref{eq:graphical2}) means that if this straight line is above the graph of $s_{\rm MF}(\varepsilon')$ for all $\varepsilon'>\varepsilon$,
then $s(\varepsilon)=s_{\rm MF}(\varepsilon)$.

\begin{figure}[tbh]
\begin{center}
\includegraphics[clip,width=7cm]{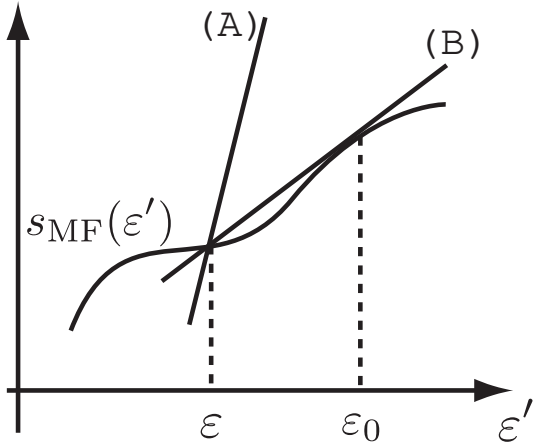}
\caption{An illustration of the upper bound of the entropy (\ref{eq:s_upper}).
We consider a straight line which passes through a point $(\varepsilon,s_{\rm MF}(\varepsilon))$ 
and has the slope $\beta_{\rm conv}/U_{\rm max}$.
(A) the line with a small $U_{\rm max}$, and (B) the line with a large $U_{\rm max}$. 
In (B), the straight line touches the graph of the entropy at $\varepsilon'=\varepsilon_0$.}
\label{fig:upper}
\end{center}
\end{figure}

In Fig.~\ref{fig:upper}, we demonstrate this aspect.
Let us imagine that we change $U_{\rm max}$ from zero to unity.
When $U_{\rm max}$ is small, the slope $\beta_{\rm conv}/U_{\rm max}$ of the straight line is large.
In this case, the straight line is above the graph of the entropy for all $\varepsilon'>\varepsilon$.
This situation corresponds to (A) of Fig.~\ref{fig:upper}.
As $U_{\rm max}$ increases, the slope decreases, and at last the straight line touches the graph of the entropy and
the RHS of (\ref{eq:s_upper}) becomes maximum at $\varepsilon'=\varepsilon_0\neq\varepsilon$, which is described by (B) of Fig.~\ref{fig:upper}.

If the condition
\beq
\frac{\beta_{\rm conv}}{U_{\rm max}}\geq \bar{\beta}
\label{eq:inflection}
\eeq
is satisfied, the inequality~(\ref{eq:graphical2}) holds for any $\varepsilon$ such that $\varepsilon_l\leq\varepsilon\leq\varepsilon_h$.
In this case, $s(\varepsilon)=s_{\rm MF}(\varepsilon)$ holds for all $\varepsilon$.
Hence, the inequality (\ref{eq:inflection}) gives a sufficient condition for the exactness of the mean-field theory.
In Sec.~\ref{sec:example}, we will explicitly examine the upper bound of the entropy (\ref{eq:s_upper}) in concrete models.

\subsection{Derivation of Eq.~(\ref{eq:s_upper})}
\label{sec:derivation}

Here, we present the derivation of Eq.~(\ref{eq:s_upper}).
Let us start with the expression (\ref{eq:s_Legendre}).
Because $\sup\inf(\cdot)\leq\inf\sup(\cdot)$,
\begin{align}
s(\varepsilon)&=\sup_{\Vec{m}}\sup_{\Vec{m}(\bm{x})}
\left\{\inf_{\beta\geq 0}
\left[\beta(\varepsilon-{\cal F}(\beta,\{ \Vec{m}(\bm{x})\})\right]:\int_{C_d}\Vec{m}(\bm{x})d^dx=\Vec{m}\right\} \nonumber \\
&\leq\sup_{\Vec{m}}\inf_{\beta\geq 0}\left\{\beta\left(\varepsilon-\inf_{\Vec{m}(\bm{x})}\left[{\cal F}(\beta,\{\Vec{m}(\bm{x})\})
:\int_{C_d}\Vec{m}(\bm{x})d^dx=\Vec{m}\right]\right)\right\} \nonumber \\
&=\sup_{\Vec{m}}\inf_{\beta\geq 0}\left[\beta(\varepsilon-f(\beta,\Vec{m})\right],
\end{align}
where $f(\beta,\Vec{m})$ is the free energy at a temperature $T=\beta^{-1}$ with a fixed value of the magnetization $\Vec{m}$.
In the first equality, we used 
$$\sup_{\Vec{m}(\bm{x})}[(\cdot)]=\sup_{\Vec{m}}\left\{
\sup_{\Vec{m}(\bm{x})}\left[(\cdot):\Vec{m}=\int_{C_d}\Vec{m}(\bm{x})d^dx\right]\right\}.$$
The free energy with a fixed magnetization satisfies the following inequality~\cite{Mori2010,Mori_instability}:
\beq
f(\beta,\Vec{m})\geq f_{\rm MF}(\beta,\Vec{m})-U_{\rm max}[f_{\rm MF}(\beta U_{\rm max},\Vec{m})-f_{\rm MF}^{**}(\beta U_{\rm max},\Vec{m})].
\eeq
Because $\inf_{\beta\geq 0}[\cdot ]\leq\inf_{0\leq\beta\leq\beta_{\rm conv}/U_{\rm max}}[\cdot ]$,
we obtain
\begin{align}
s(\varepsilon)&\leq\sup_{\Vec{m}}\inf_{0\leq\beta\leq\beta_{\rm conv}/U_{\rm max}}
\left[\beta\left\{\varepsilon-f_{\rm MF}(\beta,\Vec{m})\right.\right.\nonumber \\
&\quad\quad\left.\left. +U_{\rm max}\left[f_{\rm MF}(\beta U_{\rm max},\Vec{m})-f_{\rm MF}^{**}(\beta U_{\rm max},\Vec{m})\right]\right\}\right] \nonumber \\
&=\sup_{\Vec{m}}\inf_{0\leq\beta\leq\beta_{\rm conv}/U_{\rm max}}\left\{\beta[\varepsilon-f_{\rm MF}(\beta,\Vec{m})]\right\}.
\label{eq:derivation1}
\end{align}

We consider the case of $\varepsilon\geq u_{\rm MF}(\Vec{m})$ and the case of $\varepsilon\leq u_{\rm MF}(\Vec{m})$ separately.
First, we consider the case of $\varepsilon\geq u_{\rm MF}(\Vec{m})$.
Because $f_{\rm MF}(\beta,\Vec{m})=u_{\rm MF}(\Vec{m})-(1/\beta)\omega(\Vec{m})$,
the minimum is realized at $\beta=0$ and we obtain
\begin{align}
\sup_{\Vec{m}}\inf_{0\leq\beta\leq\beta_{\rm conv}/U_{\rm max}}
&\left[\beta(\varepsilon-f_{\rm MF}(\beta,\Vec{m})):\varepsilon\geq u_{\rm MF}(\Vec{m})\right]\nonumber \\
&=\sup_{\Vec{m}}\left[\omega(\Vec{m}):\varepsilon\geq u_{\rm MF}(\Vec{m})\right] \nonumber \\
&=s_{\rm MF}(\varepsilon).
\end{align}

Next, let us consider the case of $\varepsilon\leq u_{\rm MF}(\Vec{m})$.
In this case, the minimum is realized at $\beta=\beta_{\rm conv}/U_{\rm max}$.
Therefore,
\begin{align}
&\sup_{\Vec{m}}\inf_{0\leq\beta\leq\beta_{\rm conv}/U_{\rm max}}
\left[\beta(\varepsilon-f_{\rm MF}(\beta,\Vec{m})):\varepsilon\leq u_{\rm MF}(\Vec{m})\right]\nonumber \\
&=\sup_{\Vec{m}}\left[\frac{\beta_{\rm conv}}{U_{\rm max}}(\varepsilon-u_{\rm MF}(\Vec{m}))+\omega(\Vec{m})
:\varepsilon\leq u_{\rm MF}(\Vec{m})\right] \nonumber \\
&=\sup_{\varepsilon'\geq\varepsilon}\sup_{\Vec{m}}\left[\frac{\beta_{\rm conv}}{U_{\rm max}}(\varepsilon-\varepsilon')+\omega(\Vec{m})
:u_{\rm MF}(\Vec{m})=\varepsilon'\right] \nonumber \\
&=\sup_{\varepsilon'\geq\varepsilon}\left[\frac{\beta_{\rm conv}}{U_{\rm max}}(\varepsilon-\varepsilon')+s_{\rm MF}(\varepsilon')\right].
\label{eq:derivation2}
\end{align}
When $\varepsilon'=\varepsilon$,
$$\frac{\beta_{\rm conv}}{U_{\rm max}}(\varepsilon-\varepsilon')+s_{\rm MF}(\varepsilon')=s_{\rm MF}(\varepsilon).$$
Therefore,
\beq
\sup_{\varepsilon'\geq\varepsilon}\left[\frac{\beta_{\rm conv}}{U_{\rm max}}(\varepsilon-\varepsilon')+s_{\rm MF}(\varepsilon')\right]
\geq s_{\rm MF}(\varepsilon).
\eeq
From Eq.~(\ref{eq:derivation1}),
\begin{align}
s(\varepsilon)&\leq\max\left\{ s_{\rm MF}(\varepsilon),\sup_{\varepsilon'\geq\varepsilon}
\left[\frac{\beta_{\rm conv}}{U_{\rm max}}(\varepsilon-\varepsilon')+s_{\rm MF}(\varepsilon')\right]\right\} \nonumber \\
&=\sup_{\varepsilon'\geq\varepsilon}\left[\frac{\beta_{\rm conv}}{U_{\rm max}}(\varepsilon-\varepsilon')+s_{\rm MF}(\varepsilon')\right].
\end{align}
In this way, we obtain Eq.~(\ref{eq:s_upper}).

Finally we comment on the upper bound (\ref{eq:s_upper}).
If we consider the van der Waals limit and put $U_{\rm max}=1$ in Eq.~(\ref{eq:s_upper}),
this upper bound exceeds $s_{\rm MF}^{**}(\varepsilon)$ which is a ``trivial'' upper bound of $s(\varepsilon)$.
This fact indicates that the upper bound (\ref{eq:s_upper}) is not the optimal estimation of the entropy for large $U_{\rm max}$.
However, this upper bound is enough to lead us to conclude that the mean-field theory is exact for small but nonzero $U_{\rm max}$
such that $\beta_{\rm conv}/U_{\rm max}\geq\bar{\beta}$.

\section{Application to generalized Potts models}
\label{sec:example}

In this section, we apply our result on the necessary and sufficient conditions for the exactness of the mean-field theory
to two examples, the $\alpha$-Potts model with annealed vacancies and the $\alpha$-Potts model with the invisible states.

\subsection{$\alpha$-Potts model with annealed vacancies}

The Hamiltonian of the $\alpha$-Potts model with annealed vacancies is given by
\beq
H=-\frac{1}{2}\sum_{ij}^N\frac{\kappa_{\alpha}}{r_{ij}^\alpha}\delta_{s_i,s_j}(1-\delta_{s_i,0})-D\sum_i^N\delta_{s_i,0},
\label{eq:H_vac}
\eeq
where $s_i\in\{ 0,1,2,\dots,q\}$ and $r_{ij}\equiv |\bm{r}_i-\bm{r}_j|$.
The constant $\kappa_{\alpha}$ is the normalization constant,
$$\kappa_{\alpha}=\frac{1}{\sum_{j(\neq i)}\frac{1}{r_{ij}^{\alpha}}},$$
which is independent of a specific site $i$ because of periodic boundary conditions.
This model is reduced to the standard $q$-state Potts model for $D=-\infty$.

Now, we consider the case of $q=2$.
If we consider the following correspondence
\beq
\Vec{\sigma}_i
=\begin{pmatrix}
(\Vec{\sigma}_i)_1 \\ (\Vec{\sigma}_i)_2
\end{pmatrix}
\equiv
\begin{pmatrix}
\delta_{s_i,1} \\
\delta_{s_i,2}
\end{pmatrix},
\eeq
the Hamiltonian (\ref{eq:H_vac}) is expressed as
\beq
H=-\frac{1}{2}\sum_{ij}^N\frac{\kappa_{\alpha}}{r_{ij}^{\alpha}}\Vec{\sigma}_i\cdot\Vec{\sigma}_j-D\sum_i^N(\Vec{\sigma}_i)_0,
\label{eq:H_vac2}
\eeq
where $(\Vec{\sigma}_i)_0\equiv 1-(\Vec{\sigma}_i)_1-(\Vec{\sigma}_i)_2$.
Correspondence between Eq.~(\ref{eq:H}) and Eq.~(\ref{eq:H_vac2}) is the following:
\beq
K(\bm{r})=\frac{\kappa_{\alpha}}{r^{\alpha}},\quad \bm{J}=\begin{pmatrix}1&0\\ 0&1\end{pmatrix}.
\eeq
The magnetization is given by
\beq
\Vec{m}=\frac{1}{N}\sum_i^N\Vec{\sigma}_i\equiv \begin{pmatrix}m_1\\ m_2\end{pmatrix}.
\eeq
Here, we define $m_0=1-m_1-m_2=\sum_i^N\delta_{s_i,0}/N$.
The Hamiltonian of the mean-field model and the rate function are, respectively, expressed as
\beq
H_{\rm MF}=N\left[-\frac{1}{2}(m_1^2+m_2^2)-Dm_0\right]
\eeq
and
\beq
\omega(\Vec{m})=-\sum_{a=0}^2m_a\log m_a.
\eeq

The bordered Hessian matrix ${\cal H}_{\rm max}$ is given by the following $2\times 2$ matrix (see Eq.~(\ref{eq:BH_max})):
\beq
{\cal H}_{\rm max}=\begin{pmatrix}
\beta U_{\rm max}-\left(\frac{1}{m_0}+\frac{1}{m_1}\right) & -\frac{1}{m_0} \\
-\frac{1}{m_0} & \beta U_{\rm max}-\left(\frac{1}{m_0}+\frac{1}{m_2}\right)
\end{pmatrix}.
\eeq
Therefore, the maximum eigenvalue of ${\cal H}_{\rm max}$ at the energy $\varepsilon$, which is denoted by $\lambda(\varepsilon)$, is given by
\beq
\lambda(\varepsilon)=\beta U_{\rm max}+\frac{1}{2}\left[-\left(\frac{2}{m_0^*}+\frac{1}{m_1^*}+\frac{1}{m_2^*}\right)
+\rt{\left(\frac{1}{m_1^*}-\frac{1}{m_2^*}\right)+\frac{4}{m_0^{*2}}}\right],
\label{eq:lambda_vac}
\eeq
where $\Vec{m}^*(\varepsilon)$ is the equilibrium magnetization of the mean-field model at the energy $\varepsilon$.

Next we derive $\beta_{\rm conv}$ of this model.
We define $x(\beta,\Vec{m})$ as
\beq
x(\beta,\Vec{m})\equiv \text{maximum eigenvalue of the matrix} \frac{\d^2f_{\rm MF}}{\d\Vec{m}^2}(\beta,\Vec{m}).
\eeq
From the definition, $\beta_{\rm conv}$ is expressed as
\beq
\beta_{\rm conv}=\sup\left[\beta>0: \inf_{\Vec{m}}x(\beta,\Vec{m})>0\right].
\eeq
Therefore, it is necessary to compute $\inf_{\Vec{m}}x(\beta,\Vec{m})$.
The explicit form of $x(\beta,\Vec{m})$ is 
\beq
x(\beta,\Vec{m})=-1+\frac{1}{2\beta}\left[\frac{2}{m_0}+\frac{1}{m_1}+\frac{1}{m_2}
-\rt{\left(\frac{1}{m_1}-\frac{1}{m_2}\right)+\frac{4}{m_0^{2}}}\right].
\eeq
This takes the infimum at $m_0=0, m_1=m_2=1/2$, and we have
\beq
\inf_{\Vec{m}}x(\beta,\Vec{m})=-1+\frac{2}{\beta}.
\eeq
Therefore, in this model,
\beq
\beta_{\rm conv}=2.
\eeq

We consider the system~(\ref{eq:H_vac}) on two-dimensional lattice for $\alpha=1$.
In this case, it is known that $U_{\rm max}\sim 0.3$.
First,  we calculate the maximum of $\omega(\Vec{m})$ under the condition $\varepsilon=H/N$.
It determines the equilibrium order parameter $\Vec{m}^*(\varepsilon)$, the equilibrium entropy $s_{\rm MF}(\varepsilon)=\omega(\Vec{m}^*)$
and the microcanonical temperature $T=1/(\d s_{\rm MF}/\d\varepsilon)=1/\beta_{\rm MF}(\varepsilon)$ in the mean-field model ($\alpha=0$).
The quantity $\bar{\beta}$ is obtained from the graph of $\beta_{\rm MF}(\varepsilon)$,
and $\lambda(\varepsilon)$ is also calculated from Eq.~(\ref{eq:lambda_vac}).
We calculated these quantities for various parameters $(\varepsilon,D)$.

When $D>0.5$, the phase transition does not occur.
On the other hand, when $D\lesssim 0.38$, the canonical and the microcanonical ensembles are equivalent and the result is trivial.
Therefore, we focus on values of $D$ between these two values.

\begin{figure}[tbh]
\begin{center}
\begin{tabular}{cc}
(a)&(b)\\
\includegraphics[clip,width=4.5cm]{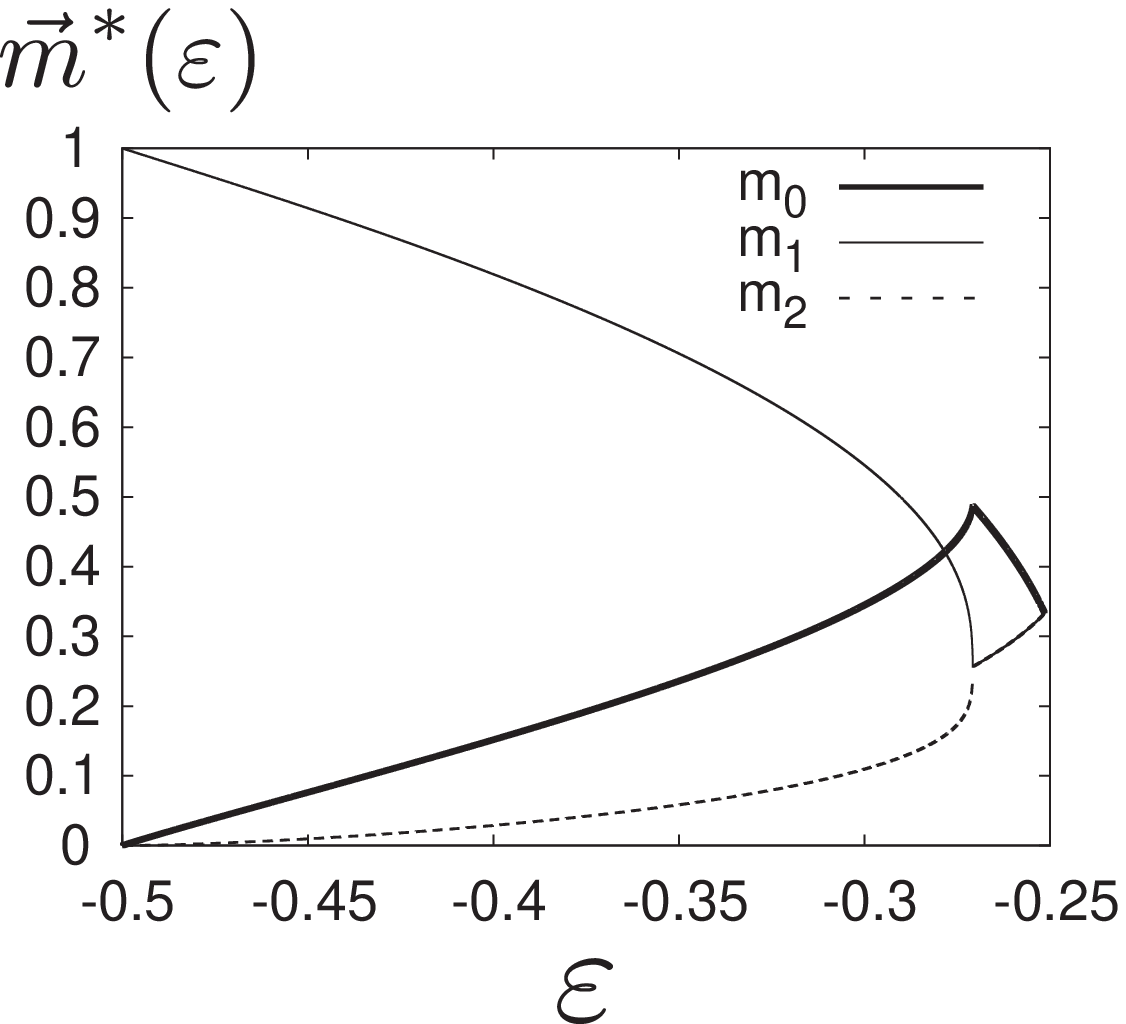}&
\includegraphics[clip,width=4.5cm]{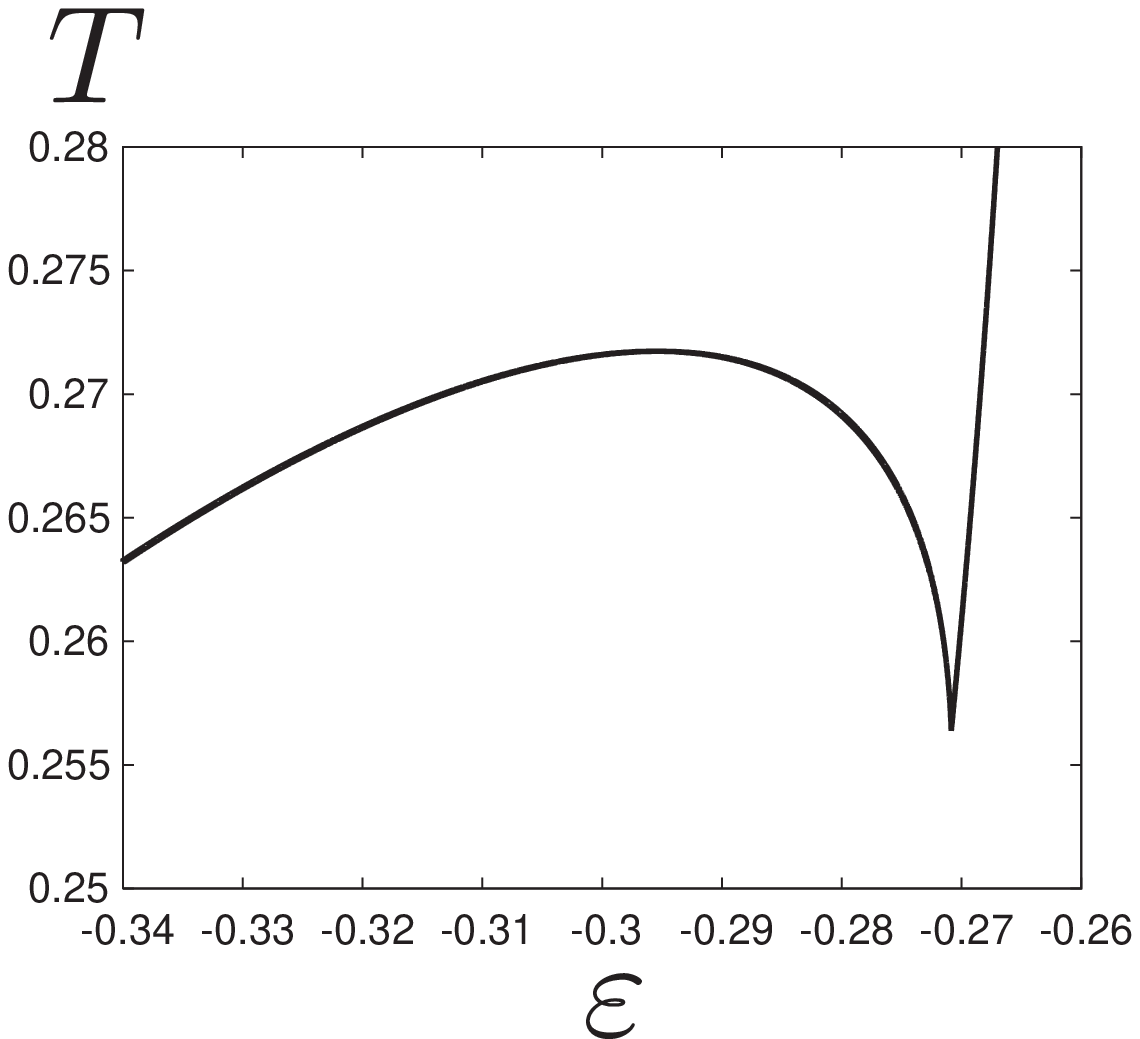}\\
(c)& \\
\includegraphics[clip,width=4.5cm]{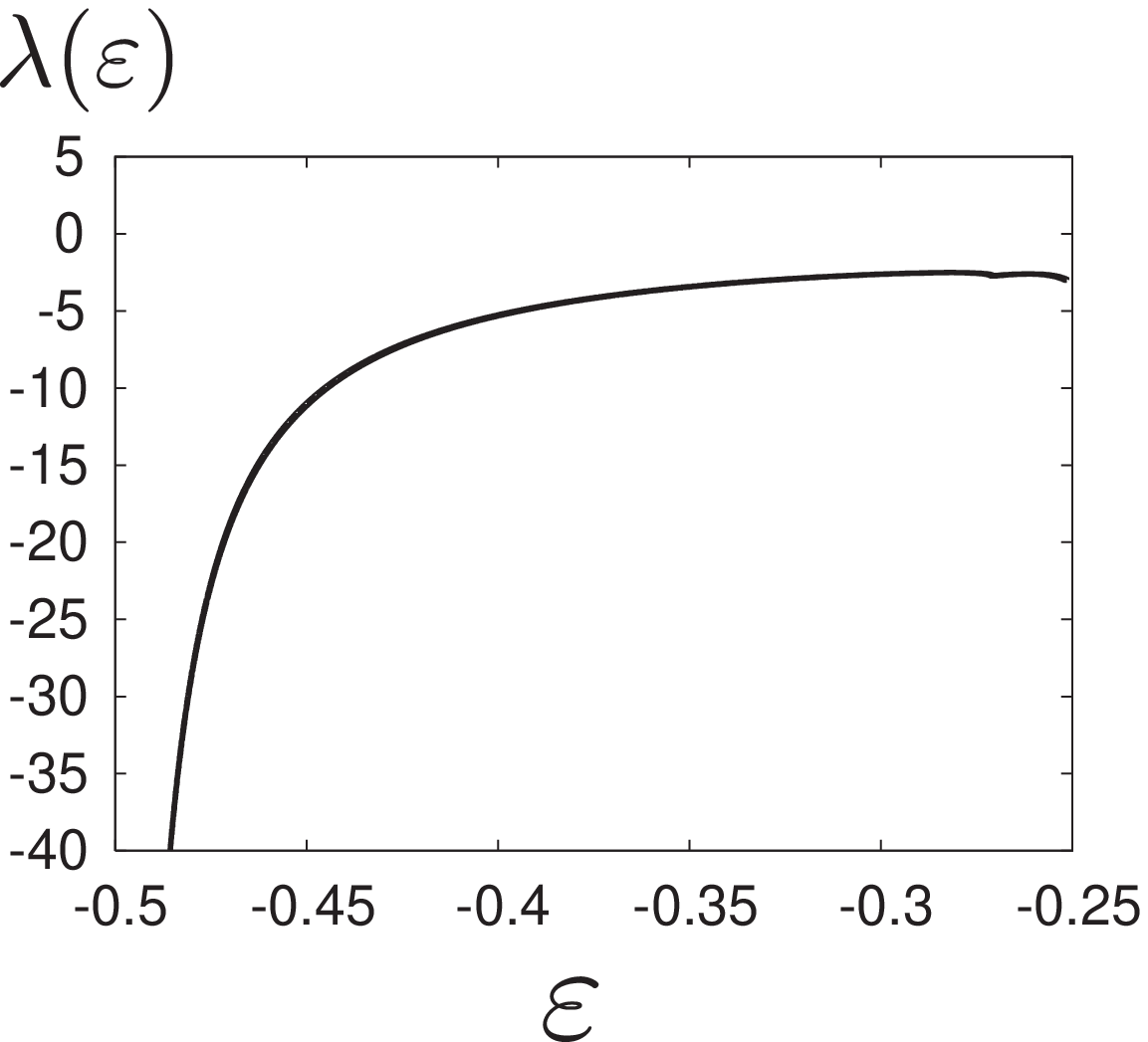}&
\end{tabular}
\caption{(a) Equilibrium magnetizations against energies in the mean-field model, 
(b) temperatures against energies in the mean-field model, 
(c) the graph of $\lambda(\varepsilon)$.
The parameter $D$ is set as $D=0.421$.}
\label{fig:vac_exm}
\end{center}
\end{figure}

Figure~\ref{fig:vac_exm} shows quantities for $D=0.421$ as functions of $\varepsilon$.
For this value of $D$, the second order phase transition occurs at $\varepsilon\approx -0.27$ (see Fig.~\ref{fig:vac_exm}(a)),
although the first order phase transition occurs in the canonical ensemble (not shown), which is a result of ensemble inequivalence. 
The negative specific heat is clearly observed in Fig.~\ref{fig:vac_exm}(b) just below the transition energy, which is a sign of ensemble inequivalence.
Figure~\ref{fig:vac_exm}(b) also shows $\bar{\beta}\approx 1/0.256\approx 3.906$, which is below $\beta_{\rm conv}/U_{\rm max}\approx 2/0.3\approx 6.7$.
Therefore, for this value of $D$, the mean-field equilibrium states are true equilibrium states for all $\varepsilon$ and mean-field theory is exact.
Figure~\ref{fig:vac_exm}(c) shows the maximum eigenvalue $\lambda(\varepsilon)$ of the bordered Hessian matrix.
In this case, $\lambda(\varepsilon)$ is negative for all $\varepsilon$, therefore the mean-field equilibrium states are always locally stable.

As far as we calculated, $\lambda(\varepsilon)$ is negative for all the parameter region of $(\varepsilon,D)$ when we put $U_{\rm max}=0.3$.
Thus the equilibrium states of the mean-field Potts model with annealed vacancies is at least locally stable for $\alpha=1$ in two dimensions.

On the other hand, we found that the sufficient condition~(\ref{eq:inflection}) is violated for $0.49\lesssim D\leq 0.5$.
Therefore, it is concluded that except for $0.49\lesssim D\leq 0.5$, two-dimensional $\alpha$-Potts model with annealed vacancies for $\alpha=1$
is equivalent to the corresponding mean-field model in the microcanonical ensemble although the canonical and the microcanonical ensembles are not equivalent.
In the region of $0.49\lesssim D\leq 0.5$, the necessary condition still holds.
Hence, the mean-field theory may be exact even in this parameter region.

\subsection{Long-range Potts model with invisible states}

Next, let us consider the long-range Potts model with the invisible states.
The mean-field version of this model is given by
\beq
H_{\rm MF}=-\frac{J}{2N}\sum_{i,j}^N\delta_{\sigma_i,\sigma_j}\sum_{a=1}^q\delta_{\sigma_i,a},
\label{eq:MF_inv}
\eeq
where the spin variable $\sigma_i$ is $\sigma_i\in\{ 1,2,\dots,q,q+1,\dots,q+R\}$.
When $R=0$, this model is identical to the standard Curie-Weiss Potts model, which is the mean-field version of the $q$-state Potts model.
The $R$ redundant states are called ``invisible states'', which do not interact with other spins (see~\cite{Tamura2010}).
The order parameter is $\Vec{m}^{\rm T}=(m_1,m_2,\dots,m_q)$ and $m_{a}=\sum_i\delta_{\sigma_i,a}/N$.
The density of invisible states is given by $m_0\equiv 1-\sum_{a=1}^qm_{a}$.
We rewrite the Hamiltonian~(\ref{eq:MF_inv}) as
\beq
H_{\rm MF}=-N\frac{J}{2}\sum_{a=1}^qm_{a}^2.
\eeq
The rate function $\omega(\Vec{m})$ is given by
\beq
\omega(\Vec{m})=-\sum_{a=0}^qm_{a}\log m_{a}-m_0\log R.
\label{eq:w_CWP}
\eeq
The Hessian matrix ${\cal H}_{\rm max}$ is given by
\beq
({\cal H}_{\rm max})_{ab}=-\left(\frac{1}{m_a}\delta_{ab}+\frac{1}{m_0}\right)+\beta JU_{\rm max}\delta_{ab}.
\label{eq:Hessian_CWP}
\eeq

The $\alpha$-Potts model with invisible states is expressed by the following Hamiltonian,
\beq
H=-\frac{1}{2}\sum_{i,j}\frac{\kappa_{\alpha}}{r_{ij}^{\alpha}}\delta_{\sigma_i,\sigma_j}\sum_{a=1}^q\delta_{\sigma_i,a}.
\label{eq:H_Potts}
\eeq
This is similar to Eq.~(\ref{eq:H_vac}).
Indeed, two models (\ref{eq:H_vac}) and (\ref{eq:H_Potts}) are equivalent in the {\it canonical ensemble},
which is observed by putting $D=T\log R$ in Eq.~(\ref{eq:H_vac}).
The only difference is that the external field $-D\sum_i\delta_{s_i,0}$ in Eq.~(\ref{eq:H_vac}) 
is replaced by the entropic term $-m_0\log R$ in Eq.~(\ref{eq:w_CWP}).
However, if we consider in the {\it microcanonical ensemble}, these two models are not equivalent as we show below.

Here, we consider the system~(\ref{eq:H_Potts}) with various values of $R$ on the two-dimensional lattice for $\alpha=1$ and $q=2$.
The analysis is almost the same as that in the previous subsection.
It is noted that $\lambda(\varepsilon)$ takes the same form as Eq.~(\ref{eq:lambda_vac}).
And $\beta_{\rm conv}$ is also the same as that of the mean-field Potts model with annealed vacancies, $\beta_{\rm conv}=2$.

\begin{figure}[tbh]
\begin{center}
\includegraphics[clip,width=7cm]{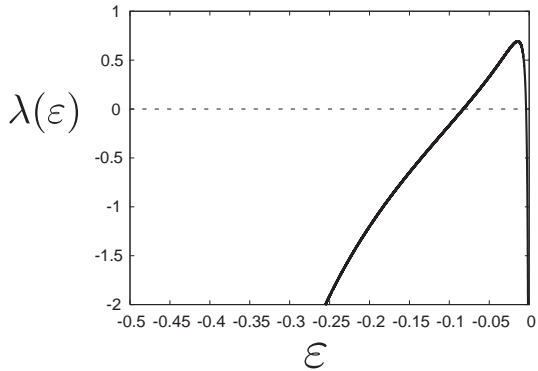}
\caption{The graph of $\lambda(\varepsilon)$ for $R=400$ and $U_{\max}=0.3$.
We can see the energy region of $\lambda(\varepsilon)>0$.
In this region, mean-field equilibrium states are locally unstable.}
\label{fig:r400}
\end{center}
\end{figure}

We calculate $\lambda(\varepsilon)$ and $\bar{\beta}$ for various $(\varepsilon, R)$.
For sufficiently large $R$ ($R\gtrsim 260$), $\lambda(\varepsilon)$ becomes positive for some $\varepsilon$, and 
the equilibrium states of the mean-field model are locally unstable in this parameter region.
This situation is demonstrated in Fig.~\ref{fig:r400}.
This is totally different from the case of the $\alpha$-Potts model with annealed vacancies.
On the other hand, from the calculation of $\bar{\beta}$, it is concluded that the mean-field theory is exact in this model at least for $R\leq 12$.
There is large difference between the necessary condition $R\lesssim 260$ and the sufficient condition $R\leq 12$.
It implies that the derived sufficient condition, $\bar{\beta}\leq \beta_{\rm conv}/U_{\rm max}$ is too strong and not optimal, 
as is mentioned in Sec.~\ref{sec:derivation}.

\begin{figure}[tbh]
\begin{center}
\begin{tabular}{cc}
(a)&(b)\\
\includegraphics[clip,width=6cm]{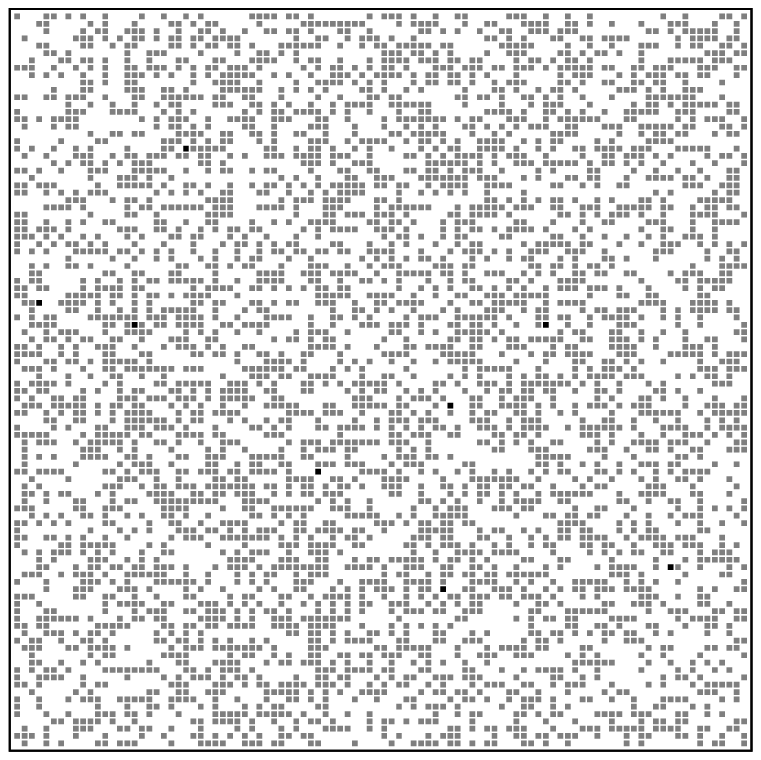}&
\includegraphics[clip,width=6cm]{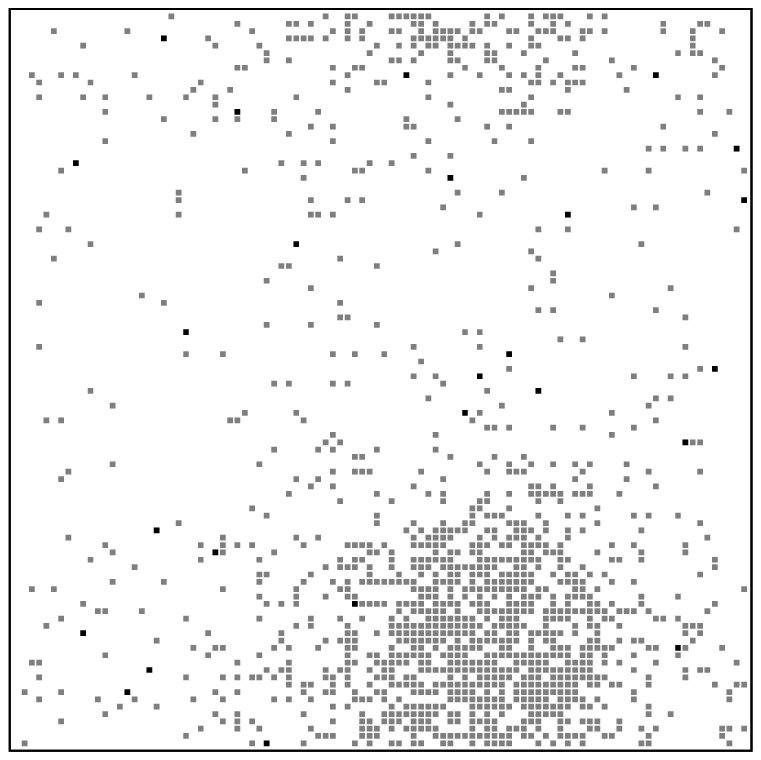}
\end{tabular}
\caption{Typical equilibrium configurations for (a) $(\varepsilon=-0.1 ,R=400)$ and (b) $(\varepsilon=-0.0125 ,R=400)$.
The system size is $L=100$.
Gray and black points correspond to $s_i=1$ and 2, respectively (but there are few black points).
The empty site corresponds to $s_i=0$.}
\label{fig:config}
\end{center}
\end{figure}

We demonstrate typical configurations of the equilibrium state calculated by the microcanonical Monte Carlo method~\cite{Creutz1983} in Fig.~\ref{fig:config}.
Clearly, the magnetization is homogeneous for $\varepsilon=-0.1$ and $R=400$ (Fig.~\ref{fig:config}(a)), 
but it is inhomogeneous for $\varepsilon=-0.0125$ and $R=400$ (Fig.~\ref{fig:config}(b)).
In Fig.~\ref{fig:config}(a) $\lambda(\varepsilon)$ is negative, and in Fig.~\ref{fig:config}(b) $\lambda(\varepsilon)$ is positive.
It is clearly observed that the homogeneous state is unstable in Fig.~\ref{fig:config}(b).
More detailed Monte Carlo analysis will be reported elsewhere.

\section{Conclusion and discussion}

We found that the exactness of the mean-field theory in long-range interacting systems holds when the canonical and the microcanonical ensembles
are equivalent in the corresponding mean-field model.
However, it does not necessarily hold in the microcanonical ensemble 
when these two ensembles are inequivalent in the mean-field model.
We derived the local stability condition for the uniform state and clarified the parameter region where the system exhibits the non-mean-field behavior.
It gives a necessary condition for the exactness of the mean-field theory.
We also derived an upper bound of the microcanonical entropy from which a sufficient condition for the exactness of the mean-field theory is obtained.
It is remarkable that these conditions are described only by the mean-field quantities and one parameter $U_{\rm max}$.
Therefore, we can judge the  validity of the exactness of the mean-field theory only by analyzing the mean-field model and calculating $U_{\rm max}$.
Because the mean-field models are numerically tractable or exactly solved, this feature of the derived conditions is preferable.
We examined these conditions for the long-range Potts model with annealed vacancies and with invisible states.

In the previous works~\cite{Mori2010,Mori_instability}, 
it was shown that long-range interacting systems can have non-mean-field equilibrium states in the canonical ensemble with a fixed value of the magnetization.
In this case, the violation of the exactness of the mean-field theory is considered to be related to the first order phase transition with varying the magnetic field
in the canonical ensemble without restriction of the magnetization value.
In this work, it has been revealed that the violation of the exactness of the mean-field theory in the microcanonical ensemble should indicate 
$s_{\rm MF}\neq s_{\rm MF}^{**}$, {\it i.e.} the ensemble inequivalence.
Inequivalence between the canonical ensemble and the microcanonical ensemble implies the presence of the first order phase transition with varying temperatures,
because the first order phase transition in the canonical ensemble makes the microcanonical entropy non-concave.

The above discussion based on the results of this work and the previous ones~\cite{Mori2010,Mori_instability} indicate an underlying general property;
{\it if a long-range interacting system exhibits a first order phase transition when an intensive parameter $x$ is varied,
then this system may have non-mean-field equilibrium states in the ensemble with a fixed extensive variable conjugate to $x$}.

In this way, the result of this work shows nontrivial properties of slowly decaying long-range interacting systems
and will promote better understanding of them.

\section*{Acknowledgements}
The author is grateful to Prof. Seiji Miyashita for useful comments and Dr. Shu Tanaka for careful reading of the manuscript.
He acknowledges JSPS for financial support (Grant No. 227835).

\appendix
\section{Bordered Hessian matrix}
\label{sec:BH}

In this Appendix, we briefly explain the method of the bordered Hessian matrix.
We want to find a maximum of a function $f(\Vec{x})$ under the condition $g(\Vec{x})=0$.
The vector $\Vec{x}$ is assumed to be a vector with $n$-components.
We introduce the Lagrange function with a Lagrange multiplier $\lambda$,
\beq
L(\lambda,\Vec{x})\equiv f(\Vec{x})-\lambda g(\Vec{x}).
\eeq
Candidates of maximum are obtained by
\beq
\frac{\d L}{\d\lambda}=0,\quad \frac{\d L}{\d\Vec{x}}=0.
\label{eq:Lagrange}
\eeq
Solutions of Eq.~(\ref{eq:Lagrange}) are denoted by $\lambda^*$ and $\Vec{x}^*$.

The solution $(\lambda^*,\Vec{x}^*)$ does not necessarily give a maximum point.
In order to judge whether the solution gives the maximum, the method of the bordered Hessian is used.
The bordered Hessian is defined as the following matrix,
\beq
{\cal H}\equiv
\begin{pmatrix}
0&-\frac{\d g}{\d x_1^*}&\cdots &-\frac{\d g}{\d x_n^*} \\
-\frac{\d g}{\d x_1^*}&\frac{\d^2L}{\d x_1^*\d x_1^*}&\cdots &\frac{\d^2L}{\d x_1^*\d x_n^*} \\
\vdots &\vdots &\ddots &\vdots \\
-\frac{\d g}{\d x_n^*}&\frac{\d^2L}{\d x_n^*\d x_1^*}&\cdots &\frac{\d^2L}{\d x_n^*\d x_n^*}
\end{pmatrix}.
\eeq
The $k$-th order minor determinant is defined as $\det {\cal H}^{(k)}$, where
\beq
{\cal H}^{(k)}\equiv
\begin{pmatrix}
0&-\frac{\d g}{\d x_1^*}&\cdots &-\frac{\d g}{\d x_{k}^*} \\
-\frac{\d g}{\d x_1^*}&\frac{\d^2L}{\d x_1^*\d x_1^*}&\cdots &\frac{\d^2L}{\d x_1^*\d x_{k}^*} \\
\vdots &\vdots &\ddots &\vdots \\
-\frac{\d g}{\d x_{k}^*}&\frac{\d^2L}{\d x_{k}^*\d x_1^*}&\cdots &\frac{\d^2L}{\d x_{k}^*\d x_{k}^*}
\end{pmatrix}.
\eeq

Whether the point $(\lambda^*,\Vec{x}^*)$ is maximum can be judged from the sign of the minor determinants of the bordered Hessian matrix.
The point $(\lambda^*,\Vec{x}^*)$ is maximum if $(-1)^k\det {\cal H}^{(k)}>0$ for all $k=2,3,\dots,n$.
See~\cite{Magnus_matrix} for more detail.


\end{document}